\documentclass[10pt,pra,twocolumn,floatfix]{revtex4}
\usepackage{amsfonts}
\usepackage{amssymb}
\usepackage{amsmath}
\usepackage[dvips]{graphicx}
\usepackage{color}

\begin{document}

\title{Manipulating the direction of Einstein-Podolsky-Rosen steering}
\author{Zhongzhong Qin$^{1,2}$}
\author{Xiaowei Deng$^{1,2}$}
\author{Caixing Tian$^{1,2}$}
\author{Meihong Wang$^{1,2}$}
\author{Xiaolong Su$^{1,2}$}
\email{suxl@sxu.edu.cn}
\author{Changde Xie$^{1,2}$}
\author{Kunchi Peng$^{1,2}$}
\affiliation{$^1$State Key Laboratory of Quantum Optics and Quantum Optics Devices,
Institute of Opto-Electronics, Shanxi University, Taiyuan 030006, People's
Republic of China\\
$^2$Collaborative Innovation Center of Extreme Optics, Shanxi University,
Taiyuan, Shanxi 030006, People's Republic of China\\
}

\begin{abstract}
Einstein-Podolsky-Rosen (EPR) steering exhibits an inherent
asymmetric feature that differs from both entanglement and Bell
nonlocality, which leads to one-way EPR steering. Although this
one-way EPR steering phenomenon has been experimentally observed,
the schemes to manipulate the direction of EPR steering have not
been investigated thoroughly. In this paper, we propose and
experimentally demonstrate three schemes to manipulate the direction
of EPR steering, either by varying the noise on one party of a
two-mode squeezed state (TMSS) or transmitting the TMSS in a noisy
channel. The dependence of the direction of EPR steering on the
noise and transmission efficiency in the quantum channel is analyzed.
The experimental results show that the direction of EPR steering of
the TMSS can be changed in the presented schemes. Our work is
helpful in understanding the fundamental asymmetry of quantum
nonlocality and has potential applications in future asymmetric
quantum information processing.
\end{abstract}

\maketitle

\section{Introduction}

Einstein-Podolsky-Rosen (EPR) steering is an intriguing phenomenon predicted
by quantum mechanics, that allows one party, say Alice, to steer the state
of a distant party, Bob, by exploiting their shared entanglement \cite%
{EPR,Schrodinger1,Schrodinger2,PrydeNatPhys}. EPR steering stands between
entanglement \cite{EPREntanglement} and Bell nonlocality \cite{Bell} in
the hierarchy of quantum correlations. Violation of the Bell inequality implies
EPR steering in both directions, and EPR steering of any direction implies
that the state is entangled. The converse implications are not true; i.e.,
entangled states are not necessarily steering states and steering does not
imply violation of the Bell inequality \cite{WisemanPRL}. In the view of
quantum information processing, EPR steering can be regarded as a verifiable
entanglement distribution by an untrusted party, while entangled states need
both parties to trust each other, and the Bell nonlocality is valid on the
premise that they distrust each other \cite{WisemanPRA}. EPR steering has
recently attracted increasing interest in quantum optics and quantum
information communities \cite{WisemanPRL,WisemanPRA,Quantifying}. For
example, it has been recently realized that EPR steering provides security
in one-sided device-independent quantum key distribution (1SDI-QKD) \cite{QKD} and
plays an operative role in channel discrimination \cite%
{ChannelDiscrimination} and teleamplification \cite{teleamplification}.
Recently, 1SDI-QKD has been experimentally implemented \cite{SchnabelOneSidedQKD,PKLamOneSidedQKD}.

\begin{figure}[tbph]
\includegraphics[width=8cm]{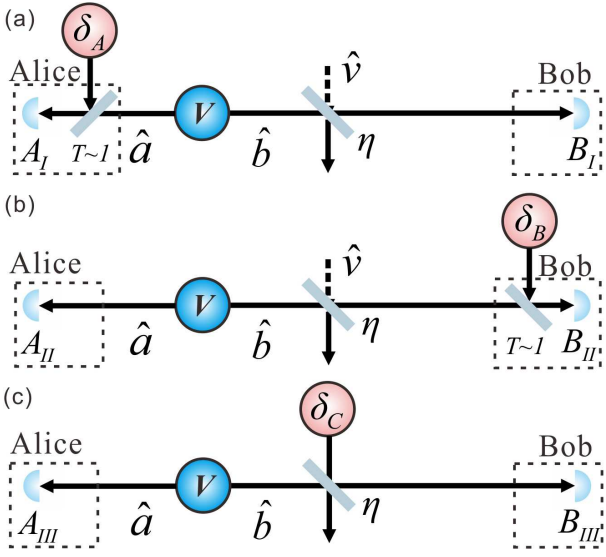}
\caption{Schematic of nontrivial manipulation schemes. (a) Manipulation scheme I.
One mode of the TMSS is transmitted over a lossy channel to Bob's station
and the loss is modeled by a beam splitter with transmission efficiency $%
\protect\eta$. Vacuum state $\hat{\protect\nu}$ is coupled into the signal
channel through another input port of the beam splitter. In this case, Alice
adds a Gaussian noise with variance $\protect\delta_{A}$ to her state by using a beam splitter
with transmission efficiency approaching unity. Alice
and Bob make measurements on the amplitude and phase quadratures $\hat{X}_{A}
$, $\hat{P}_{A}$ and $\hat{X}_{B} $, $\hat{P}_{B}$ of two remote modes by
homodyne detectors, respectively. $V$ represents the variance of one mode of
the TMSS. (b) Manipulation scheme II. Bob adds a Gaussian noise with
variance $\protect\delta_{B} $ to his state by using a beam splitter
with transmission efficiency approaching unity. (c)  Manipulation scheme III.
One mode of the TMSS is transmitted in a noisy channel with an excess noise higher than the
vacuum noise. The excess noise with variance $\protect\delta_{C}$ is coupled
to the state through the beamsplitter.}
\end{figure}

In EPR steering, Alice's ability to steer Bob's state may not be equal to
Bob's ability to steer Alice's state. There are situations where Alice can
steer Bob's state but Bob can not steer Alice's state, or vice versa, which
are referred to as one-way EPR steering \cite%
{OneWayNatPhot}. The demonstration of fundamentally asymmetric nonlocality is of
foundational significance for studying complex effects of quantum mechanics
and has potential applications in asymmetric quantum information processing.
The one-way EPR steering was first demonstrated with a two-mode squeezed
state (TMSS) \cite{OneWayNatPhot} and then extended to a multipartite system
\cite{OneWayPKLam}. Both of the above two experimental demonstrations were
performed on Gaussian states with Gaussian measurements. Other measurement
methods used to show the property of one-way EPR steering have been
theoretically constructed, including general projective measurements \cite%
{Bowles}, arbitrary finite-setting positive-operator-valued-measures (POVMs)
\cite{Quantifying}, infinite-setting POVMs \cite{Quintino}, and an infinite
number of arbitrary projective measurements \cite{Evans}. Recently, genuine
one-way EPR steering was experimentally demonstrated by two groups
independently \cite{OneWayPryde,OneWayGuo}. 

Although this one-way EPR steering phenomenon has been investigated extensively \cite{WisemanPRL,OneWayPRA,SteeringCV,Quantifying,Bowles,Quintino,Evans,ReidJOSAB,HePRL,OneWayNatPhot,OneWayPKLam,OneWayPryde,OneWayGuo,OneWayPath},
the scheme to manipulate actively the direction of EPR steering has not been
investigated thoroughly. In this paper, we propose and experimentally demonstrate three
schemes to effectively manipulate the direction of EPR steering. Two schemes are implemented by adding
noise to one party of a TMSS, and the third scheme is implemented by transmitting the TMSS
over a noisy channel. The dependence of the direction of the one-way EPR
steering upon the physical parameters is given, which offers a direct
reference for practical applications of EPR steering.

The paper is organized as follows. We present the manipulation
schemes in Sec. II. The details of the experiment are presented in Sec.
III. In Sec. IV, we present the results and some discussion.
Finally, we conclude the paper in Sec. V.

\section{Manipulation schemes}

The variance of each party in a symmetric TMSS is $V=$Cosh$2r$, where $r\in
\lbrack 0,\infty)$ is the squeezing parameter. All Gaussian properties of the
TMSS state can be determined by the covariance matrix
\begin{equation}
\sigma _{AB}=\left(
\begin{array}{cccc}
\alpha & 0 & \gamma & 0 \\
0 & \alpha & 0 & -\gamma \\
\gamma & 0 & \beta & 0 \\
0 & -\gamma & 0 & \beta%
\end{array}%
\right) =\left(
\begin{array}{cc}
\alpha I & \gamma Z \\
\gamma Z & \beta I%
\end{array}%
\right) ,
\end{equation}%
with the matrix element $\sigma _{ij}=\langle \hat{\xi}_{i}\hat{\xi}_{j}+\hat{\xi%
}_{j}\hat{\xi}_{i}\rangle /2-\langle \hat{\xi}_{i}\rangle \langle \hat{\xi}%
_{j}\rangle $, where $\hat{\xi}\equiv (\hat{X}_{A},\hat{P}_{A},\hat{X}_{B},%
\hat{P}_{B})$ is the vector of the field quadratures, including the
amplitude $(\hat{X}=\hat{a}+\hat{a}^{\dag})$ and phase $[\hat{P}=(\hat{a}-%
\hat{a}^{\dag})/i]$ quadratures of Alice's and Bob's modes. $I$ and $Z$ are
the Pauli matrices:
\begin{equation}
I=\left(
\begin{array}{cc}
1 & 0 \\
0 & 1%
\end{array}%
\right) ,\quad Z=\left(
\begin{array}{cc}
1 & 0 \\
0 & -1%
\end{array}%
\right) .
\end{equation}
Note that the submatrices $\sigma _{A}=\alpha I$ and $\sigma _{B}=\beta I$
are the covariance matrices corresponding to the states of Alice's and Bob's
subsystems, respectively. The TMSS is a symmetric state and a asymmetric state
when $\sigma _{A}=\sigma _{B}$ and $\sigma _{A}\neq \sigma _{B}$,
respectively.

EPR steering for bipartite Gaussian states of continuous variable systems
can be quantified by \cite{QuantificationPRL}
\begin{equation}
\mathcal{G}^{A\rightarrow B}(\sigma _{AB})=%
\mbox{$\max\big\{0,\,
\frac12 \ln {\frac{\det \sigma_{A}}{\det \sigma_{AB}}}\big\}$},
\end{equation}
where $\sigma _{A}$ and $\sigma _{AB}$ are the covariance matrices
corresponding to Alice's state and the TMSS state, respectively. $\mathcal{G}%
^{A\rightarrow B}(\sigma _{AB})>0$ represents that Alice has the ability to
steer Bob's state. Similarly, we have
\begin{equation}
\mathcal{G}^{B\rightarrow A}(\sigma _{AB})=%
\mbox{$\max\big\{0,\,
\frac12 \ln {\frac{\det \sigma_{B}}{\det \sigma_{AB}}}\big\}$},
\end{equation}
which represents Bob's ability to steer Alice's state, where $\sigma _{B}$
is the covariance matrix of Bob's state. From the expressions of $\mathcal{G}%
^{A\rightarrow B}(\sigma _{AB})$ and $\mathcal{G}^{B\rightarrow A}(\sigma
_{AB})$, it can be seen that Alice and Bob have the same steerability if $%
\det \sigma _{A}=\det \sigma _{B}$ is satisfied; i.e., the bipartite
Gaussian state is a symmetric state. If the state is an asymmetric state,
the steerabilities of Alice and Bob will be different.

As shown in Fig. 1(a), in manipulation scheme I, Alice adds the Gaussian
noise $\hat{N}_{A}$ with variance $\delta _{A}$ to her state. In this case,
the modes at Alice's and Bob's stations are $\hat{A}_{I}=\hat{a}+\hat{N}_{A}$
and $\hat{B}_{I}=\sqrt{\eta }\hat{b}+\sqrt{1-\eta }\hat{\nu}$, respectively.
So we have $\alpha _{I}=V+\delta _{A}$, $\beta _{I}=\eta V+(1-\eta )$, and $%
\gamma _{I}=\sqrt{\eta (V^{2}-1)}$ in the covariance matrix, where the
variances of the amplitude and phase quadratures for the thermal and vacuum
states are $\Delta ^{2}(\hat{X}_{N_{A}})=\Delta ^{2}(\hat{P}_{N_{A}})=\delta
_{A}$ and $\Delta ^{2}(\hat{X}_{\nu })=\Delta ^{2}(\hat{P}_{\nu })=1,$
respectively.

As shown in Fig. 1(b), in manipulation scheme II, Bob adds the Gaussian
noise $\hat{N}_{B}$ with variance $\delta _{B}$ to his state after the
transmission of mode $\hat{b}$ over a lossy channel. In this case, the modes
at Alice's and Bob's stations are $\hat{A}_{II}=\hat{a}$ and $\hat{B}_{II}=%
\sqrt{\eta }\hat{b}+\sqrt{1-\eta }\hat{\nu}+\hat{N}_{B}$, respectively. So
we have $\alpha _{II}=V$, $\beta _{II}=\eta V+(1-\eta )+\delta _{B}$, and $%
\gamma _{II}=\sqrt{\eta (V^{2}-1)}$ in the covariance matrix.

In EPR steering, Alice can steer Bob's state, which means that Alice can infer
Bob's state, and vice versa. In the case with unit transmission
efficiency, if noises are added to Alice's state, the uncertainty of Alice's
state predicted by Bob will increase, and thus the difficulty for Bob to steer
Alice's state will also increase and Bob will lose the ability to steer Alice's
state when the added noise is high enough. However, in this case Alice still
can steer Bob's state because the noise is added by herself and she can easily
remove the influence of the noise as needed. The analysis is also
appropriate to manipulation scheme II.

The one-way EPR steering has been demonstrated in a lossy channel previously
\cite{OneWayNatPhot,OneWayPKLam}. Besides the lossy channel, there is also
another kind of quantum channel, i.e., the noisy channel. In a lossy but
noiseless quantum channel, the noise induced by loss
is nothing but the vacuum noise (corresponding to a zero-temperature
environment). While in a noisy channel,  excess noise higher than the
vacuum noise exists \cite{RMP}. It has been shown that the excess noise in the quantum channel will limit the transmission distance of quantum key
distribution with continuous variables \cite{RMP}. As shown in Fig. 1(c), in manipulation scheme III, we consider
EPR steering of a TMSS in a noisy channel. After mode $\hat{b}$ is
transmitted through the noisy channel with excess noise $\hat{N}_{C}$, the
modes at Alice's and Bob's stations are $\hat{A}_{III}=\hat{a}$ and $\hat{B}%
_{III}=\sqrt{\eta }\hat{b}+\sqrt{1-\eta }(\hat{N}_{C}+\hat{\nu})$,
respectively. The variances of the amplitude and phase quadratures of excess
noise are $\Delta ^{2}(\hat{X}_{N_{C}})=\Delta ^{2}(\hat{P}_{N_{C}})=\delta
_{C}$. $\delta _{C}=0$ means that there is no the excess noise, and only
loss exists in the channel. $\delta _{C}>0$ means that there
excess noise exists in the channel. So we have $\alpha _{III}=V$, $\beta _{III}=\eta
V+(1-\eta )(\delta _{C}+1)$, and $\gamma _{III}=\sqrt{\eta (V^{2}-1)}$ in the
covariance matrix.

\begin{figure}[t]
\includegraphics[width=8.5cm]{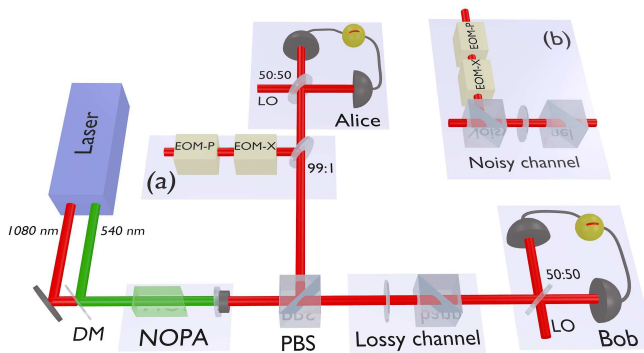}
\caption{Schematic of the experimental setup. Laser beams at 540 and 1080 nm
from a dual-wavelength laser are used as the pump and seed beams of a NOPA,
respectively. Homodyne detections are performed at Alice's and
Bob's stations, respectively. A half-wave plate and PBS are used to adjust
the transmission efficiency of the channel to Bob's station. Noise at
Alice's (or Bob's) station is added by mixing the signal with an auxiliary
beam on a high transmissivity beamsplitter (99:1), where the Gaussian white
noises are modulated on the auxiliary beam by EOMs [Inset (a)]. The
noisy channel is modeled by combining the signal channel and another
auxiliary beam modulated by EOMs at a PBS followed by another half-wave
plate and a PBS [Inset (b)]. DM, dichroic mirror; NOPA, nondegenerate
optical parametric amplifier; LO, local oscillator; EOM, Electro-optic
modulator.}
\end{figure}

\begin{figure*}[tbph]
\includegraphics[width=16cm]{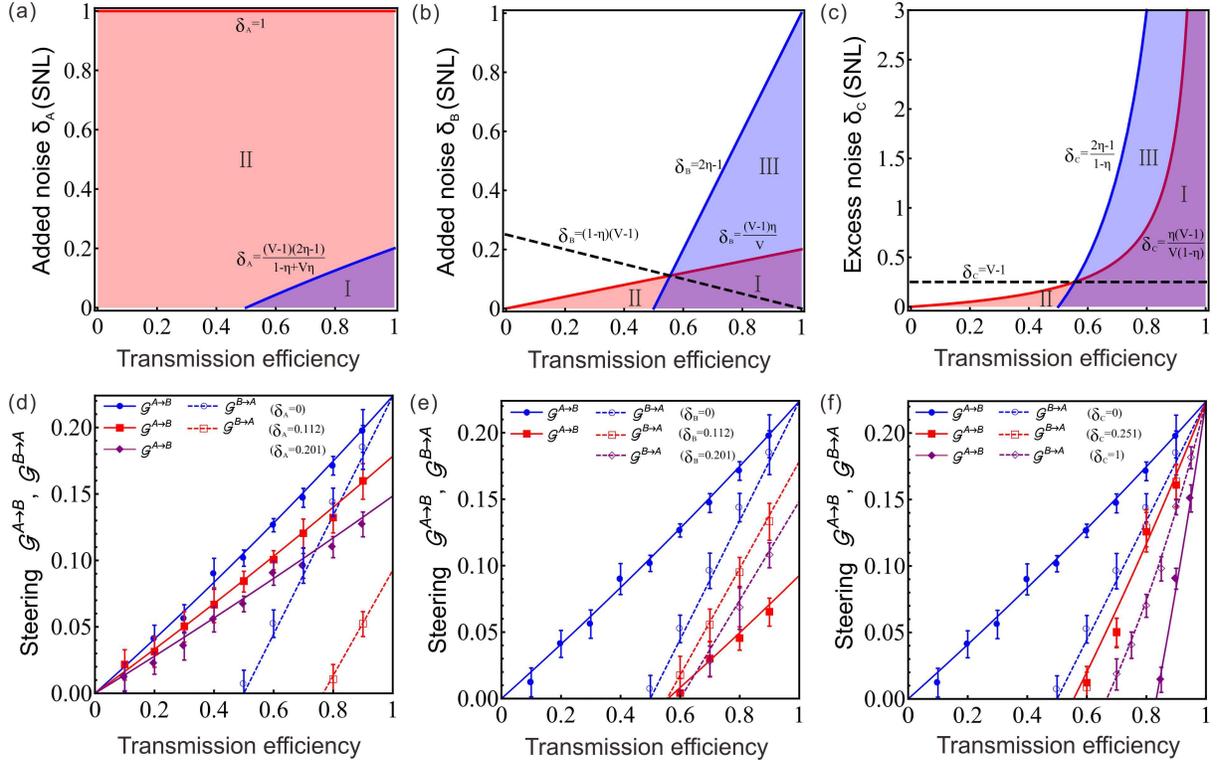}
\caption{Calculated parameter dependencies of EPR steering direction
and experimental results. (a)-(c) Different regions of EPR steering
for a TMSS with noise added to Alice's state, Bob's state, and the
channel to Bob's station, respectively. The unit of noise is
the shot noise level (SNL). Red curves and blue curves represent
boundaries for $\mathcal{G}^{A\rightarrow
B}(\protect\sigma _{AB})\geq 0$ and $\mathcal{G}^{B\rightarrow A}(\protect%
\sigma _{AB})\geq 0$ in three different cases, respectively. Region I
corresponds to two-way steering $A\leftrightarrow B$, while regions II and
III correspond to one-way EPR steering $A\rightarrow B$ and $B\rightarrow A$%
, respectively. The black dashed line $\protect\delta_{B}=(1-\protect\eta%
)(V-1)$ in panel (b) and $\protect\delta_{C}=V-1$ in panel (c) represent the condition
for $\det \protect\sigma _{A}=\det \protect\sigma _{B}$, i.e., $\mathcal{G}%
^{A\rightarrow B}=\mathcal{G}^{B\rightarrow A}$. (d)-(f) Quantum
steering of a TMSS as a function of transmission efficiency
$\protect\eta $, with different amounts of noise added to Alice's
state, Bob's state, and the channel to Bob's station, respectively.
Solid lines and dashed lines
show theoretical predictions of $\mathcal{G}^{A\rightarrow B}$ and $\mathcal{%
G}^{B\rightarrow A}$, respectively. Blue lines in the three cases represent
the situation in which there is no noise added ($\protect\delta_{A,B,C}=0$)
but only the loss exists. Red lines and purple lines in panel (d) [panel (e)] represent $%
\protect\delta_{A}=0.112$ [$\protect\delta_{B}=0.112$] and $\protect\delta%
_{A}=0.201$ [$\protect\delta_{B}=0.201$], respectively. Red lines and purple
lines in (f) represent $\protect\delta_{C}=0.251$ and $\protect\delta_{C}=1$%
, respectively. The variance of the original TMSS is chosen to be $V=1.251$
in theoretical predictions, corresponding to -3 dB squeezing. Error bars of
experimental data represent 1 standard deviation and are obtained
based on the statistics of the measured data.}
\end{figure*}

\section{Details of the experiment}

Figure 2 shows the schematic of the experimental setup. The nondegenerate
optical parametric amplifier (NOPA) consists of an $\alpha$-cut type-II KTP
crystal and a concave mirror. The front face of the KTP crystal is coated to be used
as the input coupler, and the concave mirror serves as the output coupler
\cite{SuNC}. The transmissivities of the input coupler at 540 and 1080 nm
are $21.2\%$ and $0.04\%$, respectively. The transmissivities of the output
coupler at 540 nm and 1080 nm are $0.5\%$ and $12.5\%$, respectively. The
end face of the KTP crystal is cut to $1^{o}$ along the $y$-$z$ plane of the crystal
and is antireflection coated for both 1080  and 540 nm, so that triple
resonance of the pump and the two subharmonic modes can be realized by tuning
the temperature and the position of the KTP crystal \cite{YYZhouOE}. A
nearly pure -3 dB TMSS is generated by NOPA.

In manipulation scheme I, Alice adds Gaussian noise to her state. The other mode of the TMSS is transmitted to Bob
through a lossy channel. In the demonstration of manipulation scheme II, the
noise addition part [Fig. 2, inset (a)] is moved to Bob's state. In the demonstration of the manipulation scheme III, the lossy
channel is replaced by a noisy channel [Fig. 2, inset (b)].

\section{Results and discussions}

Figures 3(a), 3(b), and 3(c) show the EPR steering regions for $\mathcal{G}%
^{A\rightarrow B}(\sigma _{AB})\geq 0$ and
$\mathcal{G}^{B\rightarrow A}(\sigma _{AB})\geq 0$ parametrized by
the transmission efficiency $\eta $ and the noise ($\delta_{A}$,
$\delta_{B}$, and $\delta_{C}$) corresponding to manipulation
schemes I, II, and III, respectively. As shown in Fig. 3(a),  two-way
EPR steering (region I) can be turned to full one-way EPR steering
$A\rightarrow B$ (region II) as the noise added to Alice's state
$\delta_{A}$ exceeds the boundary $\delta_{A}=\frac{(V-1)(2\eta
-1)}{1-\eta +V\eta }$ (blue curve). In region II, Alice can
steer Bob's state while Bob cannot steer Alice's state. This shows
that Alice stops Bob from steering her state by adding noise to her
state (while she still can steer Bob's state). The maximum noise added to Alice's state for one-way EPR steering $%
A\rightarrow B$ is $\delta_{A}=1$.

As shown in Fig. 3(b), if Bob adds noise to his state, the two-way
EPR steering can be turned to either one-way EPR steering
$A\rightarrow B$ (region II) or $B\rightarrow A$ (region III),
depending on the transmission efficiency $\eta $ and the noise level
$\delta_{B}$
added by himself. The boundary between two-way and one-way EPR steering $%
A\rightarrow B$ ($B\rightarrow A$) is given by $\delta_{B}=2\eta -1 $ [$%
\delta_{B}=\frac{(V-1)\eta }{V}$]. The crossover point at ($\eta =\frac{V}{%
1+V}$, $\delta_{B}=\frac{V-1}{V+1}$) corresponds to the boundary of
changing the direction of one-way EPR steering. This shows that
Bob can stop Alice from steering his state by adding noise to his
state in some region.

When one mode of a TMSS is transmitted in a noisy channel, Alice's
one-way EPR steering can survive iff $\delta_{C}<V-1$ and $\eta
<\frac{V}{1+V}$ [region II in Fig. 3(c)], and Bob
can achieve the one-way EPR steering in a channel with the excess noise $%
\frac{\eta (V-1)}{V(1-\eta )}<\delta_{C}<\frac{2\eta -1}{1-\eta }$ as long
as $\eta >\frac{V}{1+V}$ [region III in Fig. 3(c)]. The boundary between
two-way and one-way EPR steering for $A\rightarrow B$ ($B\rightarrow A$) is
given by $\delta_{C}=\frac{2\eta -1}{1-\eta }$ [$\delta_{C}=\frac{\eta (V-1)%
}{V(1-\eta )}$]. The crossover point at ($\eta =\frac{V}{1+V}$, $%
\delta_{C}=V-1$) corresponds to the boundary of changing the
direction of one-way EPR steering. This shows that the
direction of EPR steering is influenced by the amount of excess noise in the quantum channel.

In order to quantify the steerabilities of Alice and Bob, the dependence of
EPR steering on transmission efficiency at different noise levels and the
corresponding experimental data are shown in Figs. 3(d), 3(e), and 3(f),
respectively. The steerabilities of Alice and Bob at different transmission
efficiencies are calculated from the experimentally measured covariance
matrices, which are obtained by homodyne measurements on Alice's and Bob's
modes \cite{Steinlechner}.

The maximum EPR steering is obtained when there is no loss ($\eta =1$) and
noise ($\delta_{A,B,C}=0$), and the steerabilities of Alice and Bob are the
same since the TMSS is a symmetric state in this case. In manipulation
scheme I [Fig. 3(d)], as $\delta_{A}$ increases to $0.112 $, both $\mathcal{G%
}^{A\rightarrow B}$ and $\mathcal{G}^{B\rightarrow A}$ decrease.
However, Alice's one-way EPR steering range increases from $0<\eta
<0.5$ to $0<\eta <0.768$. As $\delta_{A}$ increases to $0.201$,
Alice obtains one-way steerability in the whole transmission
efficiency range. This result confirms that Alice  stops Bob
from steering her state under certain conditions by making local
actions (adding noise to her state).

In manipulation scheme II [Fig. 3(e)], as $\delta_{B}$ increases to $0.112$, $\mathcal{G}%
^{B\rightarrow A}$ is always larger than $\mathcal{G}^{A\rightarrow
B}$ at different transmission efficiencies, and they decrease to
zero at the same transmission efficiency $\eta=0.556$, which means
that Alice's and Bob's EPR steering ranges are the same. As
$\delta_{B}$ increases to $0.201$, Bob obtains one-way steerability
in the range of $0.601<\eta <1$. This result confirms that Bob
 stops Alice from steering his state under
certain conditions by making local actions (adding noise to his
state).

In manipulation scheme III [Fig. 3(f)],
$\mathcal{G}^{A\rightarrow B}$ and $\mathcal{G}^{B\rightarrow A}$
overlap with $\delta_{C}=0.251$, which corresponds to the case where
$\delta_{C}=V-1$ is satisfied [the black dashed line shown in Fig.3
(c)], and they decrease to zero at the same transmission efficiency
of $\eta =0.558$. As $\delta_{C}$ increases to $1$, Bob's
steerability is always larger than that of Alice, and Bob obtains
one-way steering in the transmission efficiency range of $0.667<\eta
<0.823$. This result confirms that the direction of EPR steering is
changed in a noisy channel.

Here, we discuss the physical reason for the change of the EPR steering
direction in a Gaussian TMSS. One-way EPR steering is related to the
asymmetric property of the TMSS. From
the expression of $\mathcal{G}^{A\rightarrow B}(\sigma _{AB})$ and $\mathcal{%
G}^{B\rightarrow A}(\sigma _{AB})$, it can be clearly seen that the
conditions corresponding to EPR steering regions I, II, and III are $\det
\sigma _{A}\&\det \sigma _{B}>\det \sigma _{AB}$, $\det \sigma _{A}>\det
\sigma _{AB}>\det \sigma _{B}$, and $\det \sigma _{B}>\det \sigma _{AB}>\det
\sigma _{A}$, respectively. Two-way EPR steering can be transformed to
one-way EPR steering $A\rightarrow B$ ($B\rightarrow A$) if the asymmetry of
the state exceeds the boundary $\det \sigma _{AB}=\det \sigma _{B}$ ($\det
\sigma _{AB}=\det \sigma _{A}$) between regions I and II (regions I and III).
However, it must be pointed out that the asymmetric property of the TMSS is
only a necessary condition for one-way EPR steering. In other words, a TMSS
exhibiting one-way EPR steering must be an asymmetric state, while a TMSS
exhibiting two-way EPR steering may also be an asymmetric state.

The presented manipulation schemes can be connected with one of the
applications of EPR steering, the 1SDI-QKD scheme. The security of
1SDI-QKD depends on the direction of EPR steering. For example,
``Alice must demonstrate steering of Bob's state" if Alice's
measurement device is untrusted while Bob's is trusted \cite{QKD}.
In manipulation scheme I, Alice's manipulation can be regarded
as part of a legitimate step of a QKD protocol. In this case, Alice
gets full one-way EPR steering over the whole transmission efficiency
range ($0<\eta <1$) by adding noise to her state. Manipulation
schemes II and III can be regarded as attacks from an adversarial
party. In manipulation scheme II, Bob obtains one-way steering
ability under certain conditions by making local actions (it can
also be done by an adversarial party in the attack). The change of
EPR steering direction may lead to the change of role of the
communication parties in 1SDI-QKD, which may influence the security
of 1SDI-QKD. It has been shown that the secure transmission
distance of 1SDI-QKD is limited by excess noise in the quantum channel
\cite{PKLamOneSidedQKD}. In manipulation scheme III, by
investigating the direction of EPR steering in a noisy channel, we
show that the excess noise results in the change of the direction of
EPR steering, which provides the physical reason for limiting the secure
transmission distance of 1SDI-QKD caused by excess noise.

\section{Conclusion}

In summary, three schemes to actively manipulate the direction of
EPR steering are demonstrated. The manipulation schemes are
implemented either by varying the noise on one party of a TMSS or by transmitting the TMSS in a noisy channel.
We show that the direction of EPR steering is related to the
asymmetry of a TMSS. The change of EPR steering direction depends on
the noise level and the transmission
efficiency of the quantum channel in different manipulation schemes. The experimental results confirm
that the direction of EPR steering of the TMSS can be changed in the
presented schemes. Our work is helpful in understanding the fundamental asymmetry of EPR
steering and has potential applications in asymmetric quantum
information processing.

This research was supported by the NSFC (Grant Nos. 11522433, 61601270, and
61475092), the program of Youth Sanjin Scholar, the applied basic research
program of Shanxi province (Grant No. 201601D202006) and National Basic
Research Program of China (Grant No. 2016YFA0301402). 

Z. Qin and X. Deng contributed equally to this work.

\end{document}